\documentclass[12pt]{article}

\usepackage{graphicx}
\usepackage{dcolumn}
\usepackage{bm}

\begin{document}
\def\p {{\partial}}
\def\n {{\nu}}
\def\m {{\mu}}
\def\a {{\alpha}}
\def\bt {{\beta}}
\def\f {{\phi}}
\def\th {{\theta}}
\def\g {{\gamma}}
\def\eps {{\epsilon}}
\def\e {{\psi}}
\def\la {{\lambda}}
\def\na {{\nabla}}
\def\k {\chi}
\def\bn {\begin{eqnarray}}
\def\en {\end{eqnarray}}
\begin{center}
\Large{\bf Fractional Hamiltonian analysis of higher\\ order derivatives systems} \\
\end{center}

\begin{center}
{ Dumitru Baleanu}\footnote[1]{On leave of absence from Institute
of Space Sciences, P.O.BOX, MG-23, R 76900, Magurele-Bucharest,
Romania,E-mails: dumitru@cankaya.edu.tr, baleanu@venus.nipne.ro},
Kenan Ta\c{s}

Department of Mathematics and Computer Sciences, Faculty of Arts
and Sciences, \c{C}ankaya University- 06530, Ankara, Turkey
\end{center}

\begin{center}
{Sami I. Muslih}\\
 Department of Physics, Al-Azhar University,
Gaza, Palestine\\ and \\International Center for Theoretical
Physics,\\Trieste, Italy
\end{center}

The fractional Hamiltonian analysis of 1+1 dimensional field
theory is investigated and the fractional Ostrogradski's
formulation is obtained. The fractional path integral of both
simple harmonic oscillator with an acceleration-squares part and a
damped oscillator are analyzed. The classical results are obtained
when fractional derivatives are replaced with the integer order
derivatives.\\

{\bf I. INTRODUCTION}\\

Fractional calculus deals with the generalization of
differentiation and integration to non-integer orders. Fractional
calculus has gain importance especially during the last three
decades$^{1-5}$.
 A large body of mathematical knowledge on fractional integrals and
derivatives has been constructed. Fractional calculus, as a
natural generalization of classical calculus, has played a
significant role in engineering, science, and pure and applied
mathematics in recent years. The fractional derivatives are the
infinitesimal generators of a class of translation invariant
convolution semigroups which appear universally as attractors.

Various applications of fractional calculus are based on replacing
the time derivative in an evolution equation with a derivative of
fractional order. The results of several recent researchers
confirm that fractional derivatives seem to arise for important
mathematical reasons$^{5-21}$.

 The fractional variational
principles represents an important part of fractional calculus and
it is deeply related to the fractional quantization procedure
There are several proposed methods to obtain the fractional
Euler-Lagrange equations and the corresponding Hamiltonians
However this issue is not yet complectly clarified and it requires
more further detailed analysis.

Quantization of systems with fractional derivatives is a novel
area in theory of application of fractional differential and
integral calculus. $Schr\ddot{o}dinger$ equation was considered
with the first order time derivative modified to Caputo fractional
ones in$^{22}$. In this case the obtained Hamiltonian was found to
be non-Hermitian and non-local in time. In addition, the obtained
wave functions are not invariant under the time reversal. The
quantization of fractional Klein-Gordon field and fractional
electromagnetic potential in the Coulomb gauge and the temporal
gauge were investigated very recently in$^{23}$.

Recently, the fractional variational principles and the fractional
Euler-Lagrange were obtained$^{24-25}$.

Even more recently, the fractional constrained Lagrangian and
Hamiltonian were analyzed$^{26-27}$. The notion of the fractional
Hessian$^{27}$ was introduced and the Euler-Lagrange equations
were obtained for a Lagrangian linear in velocities$^{26}$.
Besides, the Hamiltonian equations have been obtained for systems
with linear velocities$^{28}$.
 The classical fields with fractional derivatives were investigated
 by using the fractional Lagrangian formulation and the fractional Euler-Lagrange equations were obtained  in$^{29}$.

 Non-local theories have been investigated in several physical
problems$^{30}$.
 During the last decade the non-local theories were subjected to an intense debate$^{31-35}$.
  A Hamilton formalism for non-local Lagrangians was developed in$^{34-35}$, an equivalent singular first order Lagrangian was obtained and
the corresponding Hamiltonian was pulled back on the phase space
by using the corresponding constraints$^{34}$. It was shown the
space-time non-commutative field theories are acausal and the
unitarity is lost$^{36-37}$. The fractional Lagrangians and
Hamiltonians are typical examples of non-local theories.

For these reasons the fractional quantization of field theory is
an interesting issue to be investigated.

 In this paper we analyze the fractional  Hamiltonian quantization of non-singular systems possessing higher order
 derivatives.

The plan of the paper is as follows: \\In section 2 the 1+1
classical dimensional field theory analysis of non-local theories
is briefly reviewed and the fractional generalization of
Ostrogradski's formulation is presented. In section 3 the path
integral quantization of the simple harmonic oscillator with an
acceleration-squares part is analyzed. Section 4 is dedicated to
the fractional path integral formulation of the damped oscillator.
Finally, Section 5 is dedicated to our conclusions.\\

{\bf II. FRACTIONAL FIELD THEORY}\\

{ \bf A. Classical non-local theory}\\

Let us start with an ordinary local Lagrangian depending on a
finite number of derivatives at a given time, namely
\begin{equation}
L\left(q(t), {\dot q(t)},..., q^{(n)}(t)\right).
\end{equation}

The next step is to consider a Lagrangian depending on a piece of
the trajectory $ q(t, \la)$ for $\forall \la$ belonging to an
interval [a, b]
\begin{equation}\label{l1}
L^{non}(t) = L (q(t + \la)),
\end{equation}
 where $a, b$ are real numbers. Therefore a non-local Lagrangian
 was introduced.
In this case the action function corresponding to (\ref{l1})  is
given by

\begin{equation}\label{lag}
S (q) = \int dt L^{non}(t)
\end{equation}

and the Euler-Lagrange equation corresponding of (\ref{lag}) are
given by

\begin{equation}\label{ecula}
\int dt \frac{\delta L^{non} (t)}{\delta(q(t))} =0.
\end{equation}

 Equations (\ref{ecula}) should be understood as a functional relation to be
satisfied by physical trajectories, i. e., a Lagrangian
constraint. These functional relations define a subspace $J_{R}$
of physical trajectories $J_{R}\subset J$, in the space of all
possible trajectories$^{32,34}$. The crucial point is that there
is no dynamics except the displacement inside the trajectory,
namely

\begin{equation}
q(t) \rightarrow q(t +\la).
\end{equation}

Let us introduce now the dynamical variable $ Q(t, \la)$ as
follows
\begin{equation}
Q(t,\la) = q(t + \la).
\end{equation}

 If we consider a field~ $Q(t,\la)$
instead of a trajectory $q(t)$, such that

\begin{equation}
{\dot Q(t, \la)} = {Q'(t, \la)},
\end{equation}

where $ {\dot Q}= \frac{\p Q(t,\la)}{\p t}~~ and ~~ Q'(t,\la)=
\frac{\p Q(t,\la)}{\p \la}$ we obtain a field theory in one
spatial and one time dimension, namely a $1+1$ dimensional
formulation of non-local Lagrangians$^{32,34}$.

 The coordinates and momenta are suppose to
have the following forms
\begin{equation}\label{aeee} Q(t, \la)=\sum_{m=0}^{\infty}e_{m}(\lambda)q^{(m)}(t),\\
P(t,\la)=\sum_{m=0}^{\infty}e^{m}(\la)p_{(m)}(t),\end{equation}
where
\begin{equation}
 \{q^{(n)}(t),p_{(m)}(t)\}=\delta_m^n
 \end{equation}
  and
\begin{equation}\label{auu}
e_{m}(\lambda)=\frac{\lambda^m}{m!},
e^{m}(\lambda)=(-\partial_{\lambda})^m\delta(\lambda).
\end{equation}
Therefore, the Hamiltonian for $1 +1 $ dimensional field becomes
\begin{equation}
H(t, [Q, P]) =\int d\la P(t, \la)Q'(t,\la) - \tilde{L}(t, [Q]),
\end{equation}
where $P$ denotes the canonical momentum of $Q$. The phase space
is $T*J$ together with the fundamental Poisson brackets
\begin{equation}
\{ Q(t,\la), P(t,\la')\}= \delta (\la -\la').
\end{equation}

The functional $\tilde{L}(t, [Q])$ is defined as follows
\begin{equation}\label{lel}
\tilde{L}(t, [Q])= \int d\la \delta(\la) {\cal L}(t,\la).
\end{equation}

By using (\ref{lel})  the primary constraint arises as given below
\begin{equation}
\phi(t, \la, [q, P])= P(t,\la) - \int d\sigma \chi(\la,
-\sigma)\varepsilon(t;\sigma,\la)\approx 0.
\end{equation}

Here $\varepsilon(t;\sigma,\la)$ and $\chi(\la,-\sigma)$ have the
following definition
\begin{equation}
\varepsilon(t;\sigma,\la)= \frac{\p {\cal L}(t, \sigma)} {\p
Q(t,\la)},~~\chi(\la, -\sigma)= \frac{\varepsilon(\la) -
\varepsilon(\sigma)}{2},
\end{equation}

where $\varepsilon(\la)$ is the sigma distribution. The
Euler-Lagrange equation is guaranteed by itself
\begin{equation}
{\dot \phi} \sim \psi = \int d\sigma \xi(t;\sigma, \la).
\end{equation}

{ \bf B. Fractional Ostrogradski's construction }\\

Higher -derivatives theories$^{38,39}$ appear naturally as
corrections to general relativity and cosmic strings$^{40}$.
Unconstrained higher-order derivatives possess specific features,
namely they have more degree of freedom than lower-derivative
theories and they lack a lower-energy bound. A method how to
remove all these problems was presented in$^{41}$. It was observed
that the non-local formulation translates into infinite order
Ostrogradski's formulation$^{34,35}$.

In this section, we would like to derive both the Lagrangian and
the Hamiltonian formalisms for non singular Lagrangians with
fractional order derivatives starting from the Hamiltonian
formalism of non local-theories$^{32}$. Let us consider  the
following Lagrangian  to start with

\begin{equation}
{ L}(q, t) = { L}(t, q^{\a_{m}}),
\end{equation}

where the generalized coordinates are defined as

\begin{equation}
q^{\a_{m}}= {}_a \textbf{D}_t^{\alpha_{m}}x(t),
\end{equation}
where m is a natural number.

 To
obtain the reduced phase space quantization, we start with the
infinite dimensional phase space $T*J(t)= \{Q(t, \la), P(t,
\la)\}$.

The key issue is to find an appropriate generalization of
(\ref{auu}) for the fractional case. As it was pointed out
in$^{32,34}$ the coordinates and the momenta are considered as a
Taylor series. Therefore, the first step is to generalize the
classical series to the fractional case.  A natural extension is
to use instead of factorial the Gamma function. In this way we
introduce naturally the generalized functions$^{42}$ instead of
$e_m(\lambda)$ and $e^m(\lambda)$ given by (\ref{auu}).

 As it is already known several fractional Taylor's series expansions were
developed$^{3,43}$, therefore we have to decide which one is
appropriate for our generalization. Since we are dealing with
fractional Riemann-Liouville derivatives we choose the
generalization proposed in$^{44}$, namely
 \begin{eqnarray}\label{ae} Q(t,
 \la)&=&\sum_{m=-\infty}^{\infty}e_{\a_{m}}(\la)q^{(\a_{m})}(t),\cr
P(t,\la)&=&\sum_{m=-\infty}^{\infty}e^{\a_{m}}(\la)p_{(\a_{m})}(t),\end{eqnarray}

where

\begin{equation}
e_{\a_{m}}(\la) = \frac {(\la-\la_0)^{\a_{m}}}{\Gamma(\a_{m} +
1)},e^{\a_{m}}(\la)= \textbf{D}_\la^{\alpha_{m}}\delta(\la-\la_0),
\end{equation}
and $\alpha_m=m+\alpha$, with $0\leq\alpha<1$.Here $\lambda_0$ is
a constant. The coefficients in (\ref{ae}) are new canonical
variables
\begin{equation}\label{ea}
\{q^{(\a_{m})}, p_{(\a_{m^{'}})}\}= \delta_{\a_m}^{\a_{m^{'}}}.
\end{equation}

By using (\ref{ea}) we obtain that

\begin{equation}\label{ewe}
\sum_{m=-\infty}^{\infty}e^{\a_m}(\la)e_{\a_m}(\la')=\delta(\la-\la'),
\end{equation}
and

\begin{equation}\label{eea}
\int_{-\infty}^{+\infty} d\la e^{\a_{m}}(\la)
e_{\a_{m^{'}}}(\la)=\delta_{{\a_m{^{'}}}}^{\a_m}.
\end{equation}

Therefore, $e^{\a_{m}}(\la)$ and $e_{\a_{m}}(\la)$ form an
orthonormal basis.

We stress on the fact that (\ref{ewe}) and (\ref{eea}) involve the
generalized functions and the relations have the meaning in the
sense of generalized functions approach$^{42,44}$.

The fractional Hamiltonian is now given by

\begin{equation}
H =\sum_{m=-\infty}^{\infty}p^{\a_{m}}q^{\a_{m+1}} - L (q^{0},
q^{\a_{m}}).
\end{equation}

The momenta constraints become an infinite set of constraints

\begin{equation}
\phi_{n} =p_{\a_{n}}(t) - \sum_{m=n}^{\infty}{}_t
\textbf{D}_b^{\alpha_{m-n}}\frac{\p L}{\p q^{(\a_{m +1})}(t)}=0.
\end{equation}
The fractional Euler-Lagrange equations are as follows
\begin{equation}
\sum_{l=-\infty}^{\infty}{}_t \textbf{D}_b^{\alpha_{l}}\frac{\p
L(t)}{\p q^{\a_{l }}(t)}=0.
\end{equation}

An interesting property of the fractional series proposed by
Riemann and discussed by Hardy in$^{44}$ is that when $\alpha_m$
becomes integers the usual form of Taylor series is obtained.
Therefore one should notice that for integer values of $\a_m$ we
have
\begin{equation}\label{os}
p_{\a_{m}}(t)-
\sum_{l=0}^{n-m-1}\left(-\frac{d}{dt}\right)^{l}\frac{\p L(t)}{\p
(\p_{t}^{l + m +1}q(t))}=0,
\end{equation}
which is the definition of Ostrogradski's momenta$^{38}$.

 In this case the Euler-Lagrange
 equation for original fractional derivative Lagrangian$^{26-30}$
 is given below

\begin{equation}
 \sum_{l=0}^{n}{}_t \textbf{D}_b^{\alpha_{l}}\frac{\p L(t)}{\p
q^{\a_{l }}(t)}=0.
\end{equation}

Now, from this equation, for integer values of $\a_m$ we obtain
the Euler-Lagrange equation for higher derivative
Lagrangian$^{32,34,38}$, namely

\begin{equation}\label{ll}
 \sum_{l=0}^{n}\left(-\frac{d}{dt}\right)^{l}\frac{\p
L(t)}{\p (\p_{t}^{l }q(t))}=0,
\end{equation}

 The constraints (\ref{os})  and (\ref{ll}) lead us to eliminate canonical
 pairs\\ $\{q^{\a_{l}}, p_{\a_{l}}\} (l\geq n)$.

 In this case the infinite dimensional phase space is reduced to a
 finite dimensional one. The reduced space is coordinated by $T*
 J^{n}= \{q^{\a_{l}}, p_{\a_{l}}\}$ with $l=0, 1,..., n-1$. The
 Hamiltonian in the reduced space is given by

\begin{equation}\label{hoho}
H =\sum_{m=0}^{n-1}p^{\a_{m}}q^{\a_{m+1}} - L (q^{0}, q^{\a_{m}}).
\end{equation}

One should notice that the canonical reduced phase space
Hamiltonian (\ref{hoho}) is obtained in terms of the reduce
canonical phase space coordinates $\{q^{\a_{l}}, p_{\a_{l}}\}$
with $l=0, 1,..., n-1$. In this case the path integral
quantization of filed system is given by

\begin{equation}
K=\int \prod_{m=0}^{n-1} dq^{\a_{m}}dp^{\a_{m}}e^{{i}\{\int dt
({\sum_{m=0}^{n-1}}p^{\a_{m}}q^{\a_{m+1}} - H)\}}.
\end{equation}

We observe that when $\a$ are integers, we obtain the path
integral for systems with higher order Lagrangians$^{32,45-46}$.\\

{\bf III. FRACTIONAL PATH INTEGRAL QUANTIZATION OF A SIMPLE
HARMONIC OSCILLATOR POSSESSING\\  ACCELERATION-SQUARES PART}\\

The classical Lagrangian to start with is given by$^{41}$
\begin{equation}\label{lh}
L=\frac{1}{2}(1+\epsilon^2\omega^2){\dot x}^2
-\frac{1}{2}\omega^2x^2-\frac{1}{2}\epsilon^2{\ddot x}^2.
\end{equation}

The fractional generalization of (\ref{lh})  has the following
form

\begin{equation}\label{eee}
L^{'}=\frac{1}{2}(1+\epsilon^2\omega^2)({{}_t
\textbf{D}_a^{\alpha}}x(t))^2
-\frac{1}{2}\omega^2x^2-\frac{1}{2}\epsilon^2{[ {}_t
\textbf{D}_a^{\alpha}({}_t \textbf{D}_a^{\alpha}x(t))]}^2.
\end{equation}

The independent coordinates are $x(t)$ and ${{}_t
\textbf{D}_a^{\alpha}}x(t)$ respectively. Let us denote
$p_1^{\alpha}=p_x$ and $p_2^{\alpha}=p_{({{}_t
\textbf{D}_a^{\alpha}x(t)})}$. The fractional canonical momenta
are$^{38}$

\begin{equation}
p_1^{\alpha}=\frac{\partial L}{\partial {}_t
\textbf{D}_a^{\alpha}x(t) }- {}_t
\textbf{D}_a^{\alpha}(\frac{\partial L}{\partial {}_t
\textbf{D}_a^{2\alpha}x(t) }), p_2^{\alpha}=\frac{\partial
L}{\partial {}_t \textbf{D}_a^{2\alpha}x(t)}.
\end{equation}
By making use of (\ref{eee}) we obtain the forms of the fractional
canonical momenta as given below

\begin{equation}\label{we}
p_1^{\alpha}=(1+\epsilon^2\omega^2){}_t
\textbf{D}_a^{\alpha}x(t)+\epsilon^2{}_t
\textbf{D}_a^{3\alpha}x(t),
\end{equation}
\begin{equation}\label{p1}
 p_2^{\alpha}=-\epsilon^2 {}_t
\textbf{D}_a^{2\alpha}x(t).
\end{equation}

Taking into account (\ref{we}) the fractional canonical
Hamiltonian becomes
\begin{equation}\label{hah}
H=p_1^{\alpha}{}_t \textbf{D}_a^{\alpha}x(t)+ p_2^{\alpha} {}_t
\textbf{D}_a^{2\alpha}x(t)-L,
\end{equation}
and after taking into account (\ref{eee}), (\ref{we}) and
(\ref{p1}) the fractional Hamiltonian has the form

\begin{equation}\label{hgh}
H=\frac{1}{2}[2p_1^{\alpha}{}_t
\textbf{D}_a^{\alpha}x(t)-\frac{(p_2^{\alpha})^2}{\epsilon^2}+\omega^2x^2(t)-(1+\epsilon^2\omega^2)({{}_t
\textbf{D}_a^{\alpha}}x(t))^2]
\end{equation}

By making use of (\ref{hgh}) the fractional path integral is given
by

\begin{equation}
K=\int dx d({}_t
\textbf{D}_a^{\alpha}x(t))dp_1^{\alpha}dp_2^{\alpha}e^{{i}\{\int
dt (p_1^{\alpha}x(t)+ p_2^{\alpha}{{}_t
\textbf{D}_a^{\alpha}}x(t)- H)\}}.
\end{equation}
\\

{\bf IV. FRACTIONAL PATH INTEGRAL QUANTIZATION OF DAMPED HARMONIC
OSCILLATOR}\\

The Lagrangian for this system in Ostrogradski's notations$^{38}$
 takes the form$^{9}$
\begin{equation}\label{mai}
L = \frac{1}{2} m q_{1}^{2}  + i \frac{\gamma}{2} q_{1/2}^{2} -
V(q_{0}),
\end{equation}

where \begin{equation} q^{\a_{n}} = {}_t \textbf{D}_b^{\alpha_n}x,
n=0,1,2.
\end{equation}
Here $\a_{0}=0,\a_{1}=\frac{1}{2},~~ \a_{2}=1$ and $q_{0}=x, q_{1}
={\dot x}, q_{\frac{1}{2}}={}_t \textbf{D}_b^{\frac{1}{2}}x,
q_{2}= {\ddot x}.$

 The expressions for canonical momenta are
\bn
&& p_{0}= i\gamma x_{(1/2)} + i m x_{(3/2)},\\
&& p_{1/2}= m {\dot x}. \en

By using (\ref{mai}) the classical Euler-Lagrange equation of
motion read as$^{9}$
\begin{equation}
 m {\ddot x} + \gamma {\dot x} + \frac{\p V}{\p x} =0.
\end{equation}

The canonical reduced Hamiltonian has the following expression
\begin{equation}
H= \frac{p_{1/2}^{2}}{2m} + q_{1/2}p_{0} -i \frac{\gamma}{2}
q_{1/2}^{2} + V(q_{0}).
\end{equation}

Therefore, the path integral representation for the above system
analyzed is given by

\begin{equation}\label{kk}
K= \int d\mu~\exp i\left[\int \left(q_{1}p_{1/2} -
\frac{p_{1/2}^{2}}{2m} +i \frac{\gamma}{2} q_{1/2}^{2}\\ -
V(q_{0})\right)dt\right],
\end{equation}
where $d\mu=dq_{0}~dp_{0}~ dq_{1/2}~dp_{1/2}$.

 The path integral representation for (\ref{kk}) is an integration over the
canonical phase space coordinates $(q_{0}, p_{0})$ and $(q_{1/2},
p_{1/2})$. Integrating over $p_{1/2}$ and  $p_{0}$, we obtain
\begin{equation}\label{39}
K =\int ~dq_{0}~dq_{1/2}~\exp i \int{\left(\frac{1}{2} m q_{1}^{2}
-  V(q_{0}) + i \frac{\gamma}{2} q_{1/2}^{2} \right)dt}.
\end{equation}

Equation (\ref{39}) can be put in a compact form as follows
\begin{equation}\label{ehe}
K =\int ~dq_{0}~e^{ i \int{\left(\frac{1}{2} m q_{1}^{2} -
V(q_{0})\right)dt}} ~dq_{1/2} ~e^{ i{\int ( i \frac{\gamma}{2}
q_{1/2}^{2} )dt}}.
\end{equation}
After performing an integration over $q_{1/2}$  (\ref{ehe})
becomes

\begin{equation}
K =C \int ~dq_{0}~e^{ i \int{\left(\frac{1}{2} m q_{1}^{2} -
V(q_{0})\right)dt}}dt,
\end{equation}
where C represents a constant.\\

{\bf V. SUMMARY}\\

The interest in fractional quantization appears because it
describes  both conservative systems and non conservative systems
as well. The fractional quantization of field theory is not an
easy task, especially when the fractional Hamiltonian is involved.
The fractional derivatives represent the generalization of the
classical ones and therefore some of the classical properties are
lost e.g., the fractional Leibniz rule, the chain rule  become
more complicated than the classical counterparts. On the other
hand, the fractional calculus represents an emerging field and it
describes better various phenomena from several area of science
and engineering.
 The fractional path integral formulation
deserves further investigations mainly because the fractional
generalization of the classical case is not yet complectly
understood. Namely, for a system possessing second class
constraints it is difficult to find the corresponding fractional
generalization. In addition, there is no fractional formulations
of the classical secondary or tertiary constraints because the
fractional Hamiltonian is not a constant of motion.

In this paper we generalize to the fractional case the non-local
theories in one space and one time dimensions via the infinite
Ostrogradski's formalism.The classical Taylor series involved in
this problem are convergent because of the properties of the Dirac's
delta function. Namely, the coordinates and the corresponding
momenta are defined as Taylor series and the Ostrogradski's
canonical pairs fulfill the classical Poisson's brackets commutation
relations. The generalization to the fractional case of all above
mentioned results is not straightforward because there exist many
formulations for the fractional Taylor series. However, a powerful
tool in fractional field theory is to work to the Riemann-Liouville
derivatives because of their important property of integration by
parts. Therefore, in this paper we focus on the fractional Taylor
series involving the Riemann-Liouville derivatives. We assumed that
the fractional Lagrangian density has a compact support in the
x-directions.
 In this work we have obtained the path integral quantization for
fractional generalization of a 1+1 dimensional non-local field
theory. The path integral formulation for the simple harmonic
oscillator with an acceleration-squares part as well as  for the
damped oscillator are obtained. It is worthwhile to mention that
the general expression for the path integral leads to the path
integral representation for systems with higher order
Lagrangians.\\

{\bf ACKNOWLEDGMENTS}\\

 D. B.  would like to thank  O. P. Agrawal and J. J.
Trujillo for interesting discussions. This work is partially
supported by the Scientific and Technical Research Council of
Turkey.\\

S. M. would like to thank the Abdus Salam International Center for
Theoretical Physics, Trieste, Italy, for support and hospitality
during the preliminary preparation of this work. This work was
done within the framework of the Associateship Scheme of the Abdus
Salam ICTP.\\

\small{ $^{1}$K. B. Oldham and J. Spanier, \emph{The Fractional
Calculus}(Academic Press, New-York, 1974).\\

$^{2}$K. S. Miller and B. Ross, \emph{An Introduction to the
Fractional Integrals and Derivatives-Theory and Applications}(
John Wiley and Sons Inc., New York, 1993).\\

$^{3}$S. G. Samko, A.A. Kilbas, and O.I.  Marichev,
 \emph{Fractional Integrals and Derivatives - Theory and
 Applications}(Gordon and Breach, Linghorne, P.A., 1993).\\

$^{4}$I. Podlubny, \emph{Fractional Differential
Equations}(Academic Press, San Diego CA, 1999).\\

$^{5}$R. Hilfer,  \emph{Applications of Fractional Calculus in
Physics}(World Scientific Publishing Company, Singapore, 2000).\\

$^{6}$G. Zaslavsky, \emph{Hamiltonian Chaos and Fractional
Dynamics}( Oxford University Press, Oxford, 2005).\\

 $^{7}$F. Mainardi, Yu. Luchko and G. Pagnini, Frac. Calc. Appl. Anal.
\textbf{4}, 153 (2001).\\

$^{8}$F. Mainardi, Chaos, Solitons and Fractals \textbf{7},
  1461 (1996).\\

$^{9}$F. Riewe,  Phys. Rev. E \textbf{53}, 1890 (1996).\\

$^{10}$F. Riewe, Phys. Rev. E \textbf{55}, 3581
1997.\\

 $^{11}$J. A. Tenreiro Machado, Frac. Calc. Appl. Anal.
\textbf{8}, 73 (2003).\\

$^{12}$J. A. Tenreiro Machado, Frac. Calc. Appl. Anal. \textbf{4},
47 (2001).\\

$^{13}$C. F. Lorenzo and T. T. Hartley, Nonlinear Dynamics
\textbf{38}, 23 (2004).\\

$^{14}$O. P. Agrawal, Nonlinear Dynamics \textbf{38}, 191
(2004).\\

$^{15}$M. F. Silva, J.A. Tenreiro Machado and A.M. Lopes, Robotica
\textbf{23}, 595 (2005).\\

$^{16}$O. P. Agrawal, Nonlinear Dynamics \textbf{38}, 323
(2004).\\

$^{17}$M. Klimek, Czech. J. Phys. \textbf{51}, 1348 (2001).\\

$^{18}$M. Klimek, Czech. J. Phys. \textbf{52}, 1247 (2002).\\

$^{19}$E. Rabei and T. Alhalholy, Int. J. Mod. Phys. A
\textbf{9}, 3083 (2004).\\

$^{20}$S. Kemplfle, I. Schafer, and H. Beyer, Nonlinear Dynamics
\textbf{29}, 99 (2002).\\

$^{21}$R. Nigmatullin and A. Le Mehaute, J. N. Crys. Sol.
\textbf{351}, 2888 (2005).\\

$^{22}$M. Naber, J. Math. Phys. \textbf{45}, 3339 (2004).\\

$^{23}$S. C. Lim and S. V. Muniandy, Phys. Lett. A, \textbf{324},
396 (2004).\\

$^{24}$O. P. Agrawal, J. Math. Anal. Appl. \textbf{272}, 368
(2002).\\

$^{25}$O. P. Agrawal, J. Appl. Mech. \textbf{68}, 339 (2001).\\

$^{26}$D. Baleanu and T. Avkar, Nuovo Cimento B \textbf{119}, 73
(2004).\\

$^{27}$D. Baleanu, \emph{Constrained systems and Riemann-Liouville
        fractional derivative}, Proceedings of \emph{$1^{st}$ IFAC
Workshop on Fractional Differentiation and its
Applications}( Bordeaux, France, July 19-21, 597 (2004)).\\

$^{28}$S. I. Muslih and D. Baleanu, J.  Math. Anal.Apl.
\textbf{304}, 599
 (2005).\\

$^{29}$D. Baleanu and S. I. Muslih, Physica Scripta \textbf{72},
119 (2005).\\

$^{30}$A. Pais  and G. E. Uhlenbeck, Phys. Rev. \textbf{79}, 145
(1950).\\

$^{31}$J. Gomis, K. Kamimura, T. Ramirez, and J. Gomis, Nucl.
Phys. B \textbf{696}, 263 (2004).\\

$^{32}$J. Gomis, K. Kamimura, and J. Llosa, Phys. Rev. D
\textbf{63}, 045003 (2001).\\

$^{33}$J. Gomis and T. Mehen, Nucl. Phys. B \textbf{ 591}, 265
(2000).

$^{34}$J. Llosa and J. Vives, J. Math. Phys. \textbf{35}, 2856
(1994).\\

$^{35}$K. Bering, hep-th/0007192.\\

$^{36}$N. Seiberg, L. Susskind, and T. Toumbas, J. H. Energ. Phys.
\textbf{6}, Art. No. 021 (2000).\\

$^{37}$L. Alvarez-Gaume and J.L.F.  Barbon, Int. J. Mod. Phys. A
\textbf{16}, 1123
 (2001).\\

$^{38}$
 D. M. Gitman and I. V. Tytin, \emph{Quantization of fields with
constraints}(Springer-Verlag, 1990).\\

$^{39}$V. V. Nesterenko,  J.  Phys. A- Math. Gen. \textbf{22},
1673 (1989).\\

$^{40}$N. D. Birell and P.C.W. Davies, \emph{Quantum field in
curved space}(Cambridge University press, Cambridge, England,
1982).\\

$^{41}$J. Z. Simon, Phys. Rev. D. \textbf{41}, 3720 (1990).\\

$^{42}$I. M. Gelfand and G. E. Shilov,  \emph{Generalized
Functions}, vol.I, \emph{Properties and Operators} (Accademic
Press, 1964).\\

$^{43}$J. J. Trujillo, J. Math. Anal. Appl. \textbf{231}, 255
(1999).\\

$^{44}$ G. H. Hardy,  J. London Math. Soc. \textbf{20}, 48
(1954).\\

$^{45}$S.I. Muslih,
 Mod. Phys. Lett. A \textbf{36}, 2382 (2002).\\

$^{46}$S. I. Muslih, Czech. J.Phys. \textbf{53}, 1163 (2003).}

\end{document}